\documentclass[11pt,twoside]{article}
\usepackage{asp2010,natbib,graphicx}

\resetcounters

\begin{document}

\title{The UV view of multi spin galaxies: insight from SPH simulations}
\author{Daniela~Bettoni,$^1$ Paola~Mazzei,$^1$ Antonina Marino,$^2$ 
Roberto~Rampazzo,$^1$ Giuseppe~Galletta,$^2$ and Lucio~M.~Buson$^1$
\affil{$^1$INAF - Osservatorio Astronomico di Padova}
\affil{$^2$Dipartimento di Fisica e Astronomia, Universit\'a di Padova}}

\begin{abstract}
The UV images of GALEX revealed that $\sim$30$\pm$3\% of Early Type Galaxies (ETG) 
show UV emission indicating a rejuvenation episode. In ETGs with multiple spin 
components this percentage increases at 50\%.
 We present here the characteristics of this sample and our smooth particle 
hydrodynamic (SPH) simulations with chemo-photometric implementation that provide 
dynamical and morphological information together with the spectral energy 
distribution (SED) at each evolutionary stage. We show our match of the global 
properties of two ETGs (e.g. NGC 3626 and NGC 5173). For these galaxies we can 
trace their evolutionary path. 
\end{abstract}

\section{Introduction}

In the hierarchical scenario of structures formation and evolution, galaxy 
interactions and mergers are very common. One of the most impressive tracer of the 
merging event is the phenomenon of counter-rotation (see \citet{Galletta96} for a 
review). Counter-rotation is observed when two components of a galaxy rotate with 
opposite spin. This may happen in a portion  or in the whole range of galaxy 
radius. The first examples of counter-rotation were found in ETGs  but in the later 
years this phenomenon has been found in every morphological type of galaxy: 
Ellipticals, S0s, Spirals and even in irregular galaxies. Due to the large number 
and variety of cases found, the counter-rotating galaxies on the last years 
abandoned the category of ``peculiar galaxies" to become examples of a phase of the 
evolution of many (or most) galaxies. 

 A current challenge is to disentangle the different evolutionary paths producing 
counter-rotating galaxy components. 
In particular the evolution of interstellar medium (ISM) in counter-rotating  
galaxies is quite complex since, in contrast with ``native'' gas  whose evolution 
is essentially determined by the star formation rate, the evolution of the accreted 
gas depends on many independent parameters, like: i) the impact angle, ii) the 
relative orientation of the accreted gas and galaxy- spins and iii) the abundance 
of native gas. 
In order to provide a complete, self-consistent, picture of both galaxy evolution 
and  dust properties, it is highly desirable to have a SED that entails the whole 
spectral range self-consistently with the galaxy evolution.
\section{The chemo-photometric SPH simulations}

To this purpose, we performed a large set of simulations of galaxy formation and 
evolution of isolated collapsing triaxial systems  initially composed of dark 
matter and gas, as described in \citep{Mazzei03}, together with a  large set of 
encounter simulations  starting from the same initial conditions of haloes, using 
different impact parameters, total masses and spin orientations.  Our simulations 
allow us to derive the global properties of interacting systems, in particular 
velocities  and relative abundances of all the system components, i.e. the merger 
evolution.

The general prescriptions of chemo-photometric SPH simulations and the grid of the 
impact parameters explored, are reported in several  previous papers 
\citep{Mazzei13, Bettoni12}.
All the simulations  include self--gravity of gas, stars and DM, radiative cooling, 
hydrodynamical pressure, shock heating,  viscosity, SF,  feedback both from  
evolved stars and type II SNe, and  chemical enrichment.
Simulations provide the synthetic SED, based on evolutionary population synthesis 
(EPS) models, at each  evolutionary stage, i.e. at each snapshot.  
The time-step between each snapshot is 0.037 Gyr. The SED accounts for chemical  
evolution, stellar  emission, internal extinction and re-emission by dust in a 
self-consistent way. This extends over four orders of magnitude in wavelength at 
least, i.e., from 0.06 to 1000 $\mu$m. 
Simulations self-consistently   provide morphological, dynamic and chemo-
photometric evolution. 
All the model parameters  had been tuned in previous papers  where the integrated 
properties of simulated galaxies stopped at  15 Gyr, i.e. their colors, absolute 
magnitudes,  mass to luminosity ratios,  and metallicities had been successfully 
compared with those of local galaxies \citep{CM99, Mazzei03}.

\section{The Sample}

Here we focus on two points: the (u-r) vs. $M_r$ color magnitude diagram and the 
fit of the SED. 
In the (u - r) vs. $M_r$ color  magnitude diagram (CMD) \citep{Baldry04} early-type 
quiescent galaxies populate the red sequence and  late-types,  with active star 
formation, the blue one.  
The Galaxy Evolution Explorer (GALEX) satellite showed that a surprisingly high 
fraction (15\%) of optically red SDSS ETGs exhibit strong UV excess \citep{Yi05}. 
Later \citet{donas07} and \citet{Schawinski} showed that up to 30$\pm$3\% of the 
ETGs imaged in the UV with GALEX have signatures of such rejuvenation episodes, 
even after excluding classical UV-upturn candidates.
Considering the whole sample of galaxies with counter-rotation \citep{Galletta96, 
Corsini98} the percentage of ETGs with signatures of rejuvenation episodes in the 
nucleus and in the disk, jump at 50\%. From this sample we selected a sub-
sample in a local volume of 10\,Mpc, that spans all the morphological sequence, 
from  Spirals to ETGs, with available data from UV to near-IR. Here we present the 
results for two ETGs: NGC 3626 (S0/Sa) and NGC 5173 (E0).

\begin{figure}[]
\includegraphics[width=180pt,height=180pt]{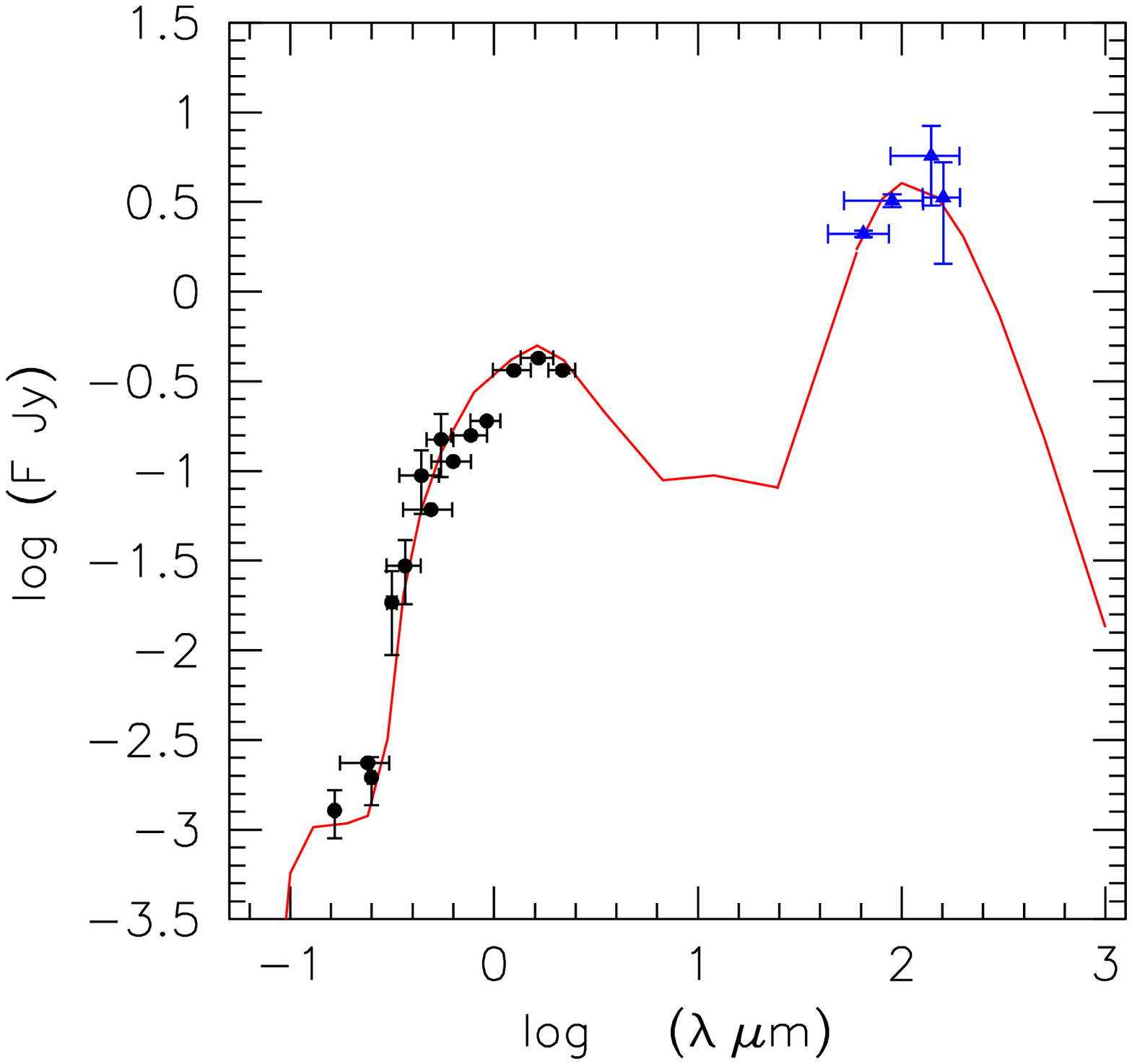}
\includegraphics[width=180pt,height=180pt]{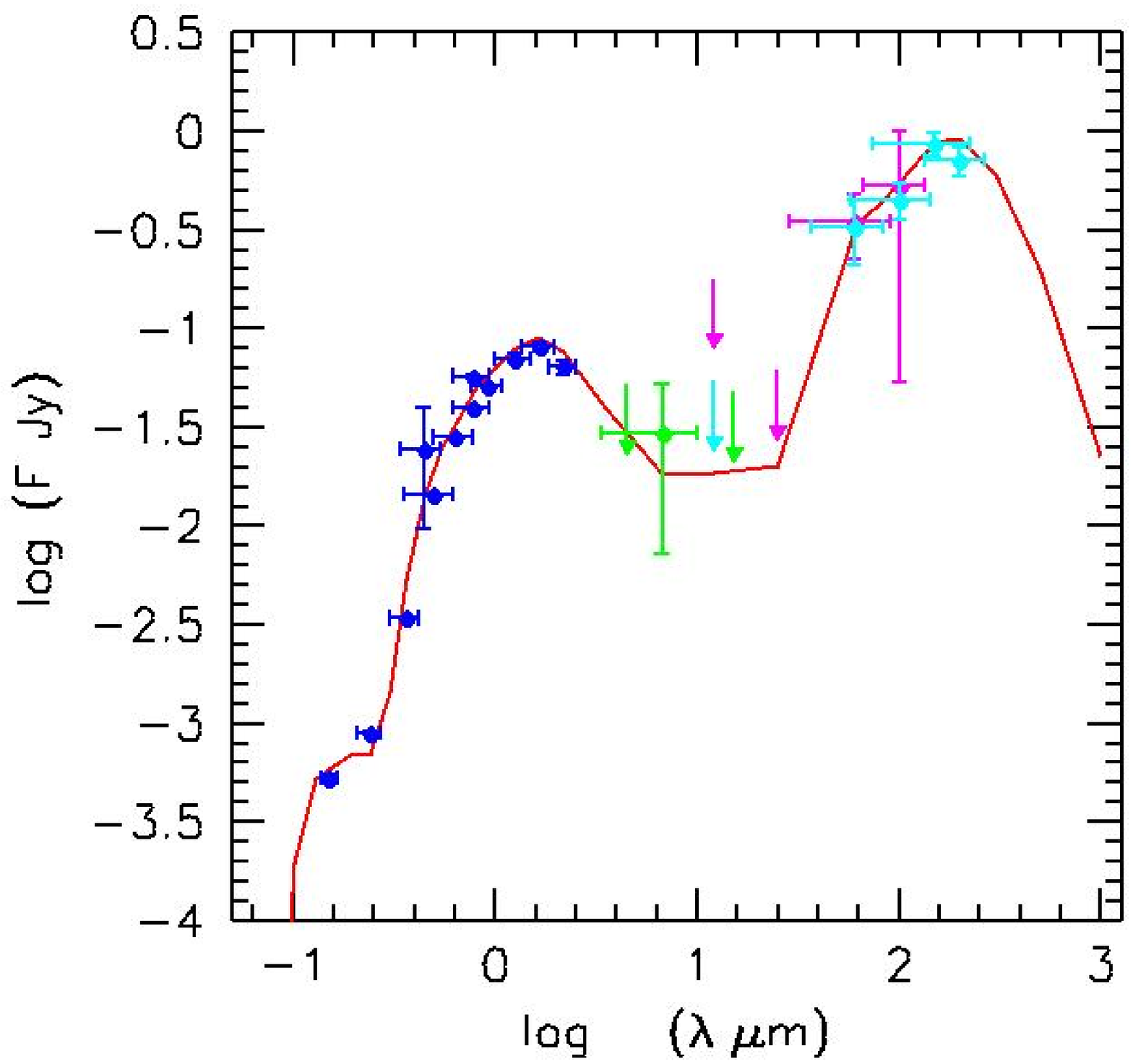}
\caption{ Data are as in \citet{Mazzei13} for 
NGC 3626 (left), with the exception of new near-UV flux from GALEX archive at MAST, whereas for NGC 5178 (right),  
green points are ISO data from \citet{Xilouris04}, magenta are IRAS data and cyan 
are from \citet{Temi04}. Solid lines are predictions from our chemo-photometric 
simulations at 13.8\,Gyr.
The FIR SED includes two dust components: warm dust (HII regions) and cold dust 
(heated by the diffuse interstellar field) both including PAH molecules (e.g. 
\citet{Bettoni12}). The warm/cold ratio is 0.23 for NGC 3626, like the average for 
Spirals \citep{Mazzei92}, and double than in average for Es for NGC 5173, i.e. 1 
instead of 0.5 \citep{Mazzei94}.
}
\label{seds}
 \end{figure}

\begin{figure}[]
\includegraphics[width=180pt,height=180pt]{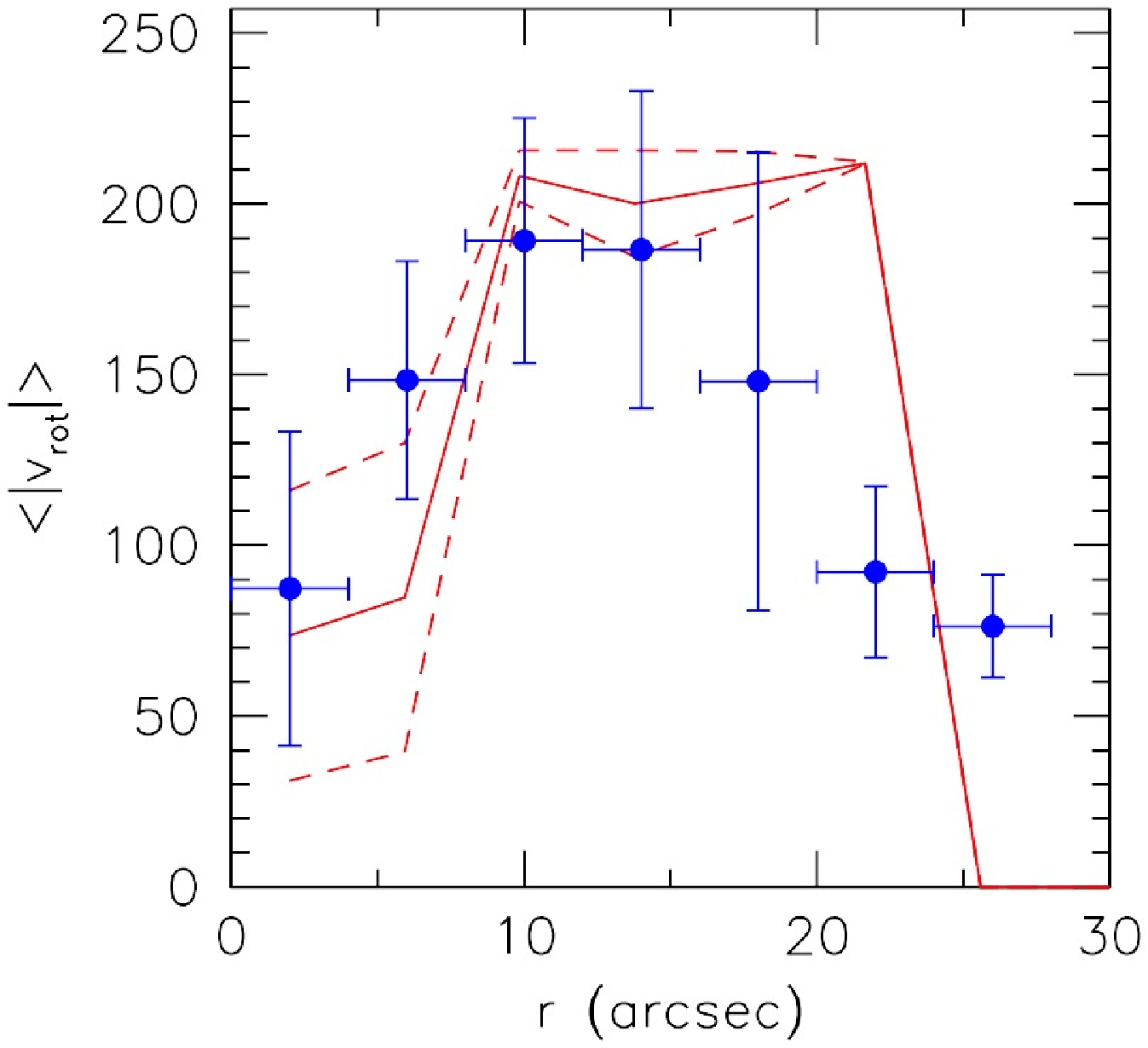}
\includegraphics[width=180pt,height=180pt]{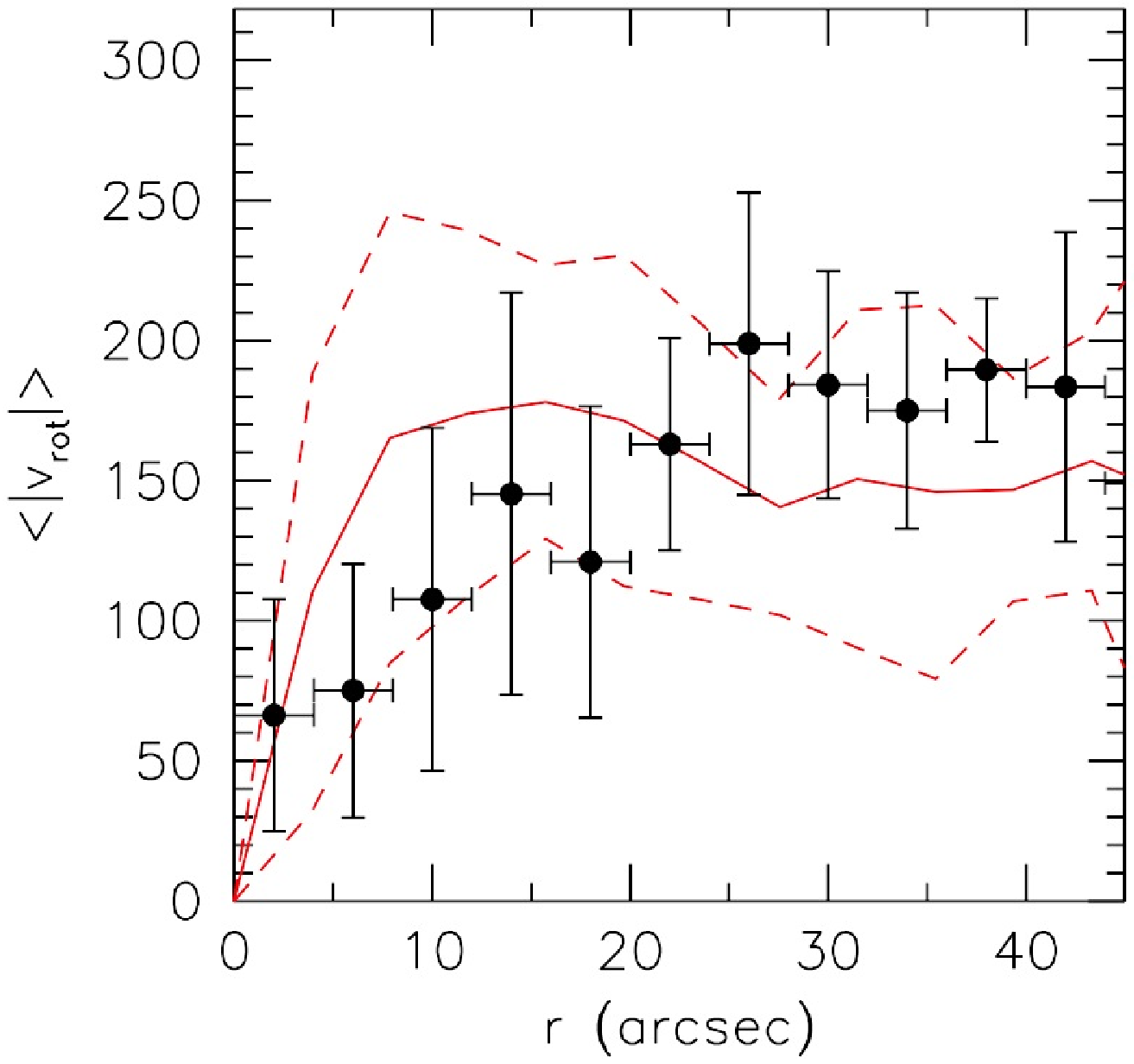}
\caption{NGC 3626 Left: Rotation curve of gas at the selected snapshot compared 
with observations \citep{Ciri, Haynes}, filled circles, both binned within 4Ë; 
vertical bars show the dispersion of observations within each bin; dashed lines 
connecting the velocities of our snapshot, show the range of model predictions. 
Right: Rotation curve of stars compared with observations \citep{Haynes}, filled 
circles; the meaning of the symbols is the same as in the left panel.}
\label{rot}
 \end{figure}  
 
 \section{Results}
{\bf{NGC 3626}}
This galaxy is the first early spiral (RSA classified it as S0) in which gas/star 
counterrotation has been found \citep{Ciri}. \cite{Garcia98} found also a disk of 
molecular gas counterrotating with respect to the stars. From our modeling 
\citep{Mazzei13}
the galaxy age results 11.5\,Gyr. The average stellar age within the effective 
radius, r$_e$, is 4.3\,Gyr and increases to 5.5\,Gyr within R$_{25}\simeq$3r$_e$. 
These age estimates become younger if stellar ages are weighted by the B-band 
luminosity: 3.4\,Gyr and 4.5\,Gyr respectively. In the inner regions, 
R$\leq$1.5\,kpc, stars are younger than 2\,Gyr, in good agreement with the findings 
of \citet{Silchenko10}. The star formation rate due to stars younger than 
0.01\,Gyr, is 2\,$M_{\odot}$ /yr, in agreement with the radio estimates given by 
\citet{Bell03}. Figure~\ref{seds}(left) compares our predictions with observations, 
including new near-UV flux released from GALEX archive; see \citet{Mazzei13} for more details. In Figure \ref{rot} we compare the observed star and gas rotation curves with the results of our simulations. \\
{\bf{NGC 5173}}
In this E0 galaxy \citet{Kannappan01} found the gas/star counterrotation and in the 
GALEX UV image it is visible a patchy FUV inner ring. The simulation which best-
fits its total SED (see Fig.\ref{seds}, right), from FUV to 200\,$\mu$m,  total B absolute magnitude, M$_B$=-
17.48$\pm$0.31 (from Hyperleda),  and  morphology at an age of 13.8\,Gyr, 
corresponds to a galaxy encounter with mass ratio 1:1, total mass 4$\times 
10^{12}$\,M$_{\odot}$ and  relative positions and velocities, r$_1$=-
r$_2$=888\,kpc,
v$_1$=-v$_2$=61\,km/s. 

\section{Conclusions}
Our simulations trace the evolution of these galaxies along the blue sequence, 
which hosts  them for about 7-8 Gyr, before they became mature and red ETGs after 
4-5 Gyr  they spent, in a significant part, in the so called green valley.   

\acknowledgements We acknowledge the partial financial support by contract ASI-INAF  
I/009/10/0, and the partial financial support by contract INAF/PRIN 2011 ``Galaxy 
Evolution with the VLT Survey Telescope (VST)".

%\bibliography{editor}
\bibliography{aspbettoni}

\end{document}